\documentclass[twocolumn]{aastex61}

\usepackage{graphicx}
\usepackage{bm}
\usepackage{tabularx}
\usepackage{lipsum}
\usepackage{gensymb}
\usepackage{multirow}
\usepackage{amsmath}
\usepackage{subfigure}
\usepackage[utf8]{inputenc}
\usepackage[section]{placeins}
\usepackage[colorlinks=true]{hyperref} 

\begin{document}

\title{Identifying Atmospheres on Rocky Exoplanets Through Inferred High Albedo}

\author{Megan Mansfield}
\affiliation{Department of the Geophysical Sciences, University of Chicago, Chicago, IL 60637, USA}
\email{meganmansfield@uchicago.edu}

\author{Edwin S. Kite}
\affiliation{Department of the Geophysical Sciences, University of Chicago, Chicago, IL 60637, USA}

\author{Renyu Hu}
\affiliation{Jet Propulsion Laboratory, California Institute of Technology, Pasadena, CA 91109, USA}

\author{Daniel D. B. Koll}
\affiliation{Department of Earth, Atmospheric, and Planetary Sciences, Massachusetts Institute of Technology, Cambridge, MA 02139, USA}

\author{Matej Malik}
\affiliation{Department of Astronomy, University of Maryland, College Park, MD 20742, USA}

\author{Jacob L.\ Bean}
\affiliation{Department of Astronomy \& Astrophysics, University of Chicago, Chicago, IL 60637, USA}

\author{Eliza M.-R. Kempton}
\affiliation{Department of Astronomy, University of Maryland, College Park, MD 20742, USA}
\affiliation{Department of Physics, Grinnell College, Grinnell, IA 50112, USA}
\date{\today}

\begin{abstract}
The upcoming launch of the \textit{James Webb Space Telescope} (\textit{JWST}) means that we will soon have the capability to characterize the atmospheres of rocky exoplanets. However, it is still unknown whether such planets orbiting close to M dwarf stars can retain their atmospheres, or whether high-energy irradiation from the star will strip the gaseous envelopes from these objects. We present a new method to detect an atmosphere on a synchronously rotating rocky exoplanet around a K/M dwarf, by using thermal emission during secondary eclipse to infer a high dayside albedo that could only be explained by bright clouds. Based on calculations for plausible surface conditions, we conclude that a high albedo could be unambiguously interpreted as a signal of an atmosphere for planets with substellar temperatures of $T_{sub}=$~410-1250~K. This range corresponds to equilibrium temperatures of $T_{eq}=$~300-880~K. We compare the inferred albedos of eight possible planet surface compositions to cloud albedo calculations. We determine that a layer of clouds with optical depths greater than $\tau=0.5$~--~$7$, would have high enough albedos to be distinguishable from a bare rock surface. This method of detecting an atmosphere on a rocky planet is complementary to existing methods for detecting atmospheres, because it provides a way to detect atmospheres with pressures below 1~bar (e.g. Mars), which are too tenuous to transport significant heat but thick enough to host high-albedo clouds.
\end{abstract}

\section{Introduction}

The \textit{Transiting Exoplanet Survey Satellite} (\textit{TESS}) is already beginning to discover small, likely rocky exoplanets around K and M dwarfs \citep[e.g.,][]{Vanderspek2018,dumusque2019,gunther2019,kostov2019,espinoza2019,luque2019,crossfield2019,winters2019}. In the near future, the \textit{James Webb Space Telescope} (\textit{JWST}) will provide the capability for atmospheric characterization of such small planets. However, it is currently unknown whether small planets around M dwarfs can retain atmospheres. The high X-ray and ultraviolet flux (``XUV'' flux) of M dwarfs may completely strip the atmospheres off small, close-in planets. This process is thought to sculpt the observed population of close-in exoplanets, dividing small planets into two categories - those with radii smaller than $\approx 1.5R_{\oplus}$, which are likely rocky cores stripped of any primordial light-element atmospheres, and those with radii larger than $\approx 2R_{\oplus}$, which retain some hydrogen and helium in their atmospheres \citep[][but see \citealt{Ginzburg2018} for an alternate explanation]{Lopez2013,Owen2013,Rogers2015,Owen2017,Fulton2018,VanEylen2018}. However, small-radius planets with periods of order 10 days can nevertheless have secondary atmospheres if the volatiles are outgassed from their interiors late in the system's history relative to the early period of high UV flux \citep{Tian2009}, are delivered by late bombardments of comets or asteroids, have high molecular weight \citep{DornHeng2018}, or are effective infrared coolants \citep{Johnstone2018}. 

One possible way to test for the presence of an atmosphere on a small planet is to look for a smaller phase curve amplitude than expected for bare rock, which for planets that are synchronously rotating would indicate the presence of an atmosphere redistributing heat to the planet's nightside \citep{Seager2009}. However, this method requires a large investment of telescope time to observe at least a half orbit, if not a full orbit, of the planet. The atmosphere could also be detected through observing features in a transmission spectrum, but clouds or hazes may obscure any features even if the planet has an atmosphere \citep[e.g.,][]{Kreidberg2014}. Emission spectroscopy can reveal atmospheric features without being limited by the presence of clouds or hazes, but it also requires a significant investment of telescope times to detect spectroscopic features, especially for cooler planets \citep{Morley2017}. In a companion paper we present a fourth method, which is to look for heat redistribution through its effect on the broadband secondary eclipse depth \citep{Koll2019}. If the planet's atmosphere is transporting heat from the dayside to the nightside, then the secondary eclipse depth will be much shallower than expected for a bare rock.

\begin{figure}
    \centering
    \includegraphics[width=\linewidth]{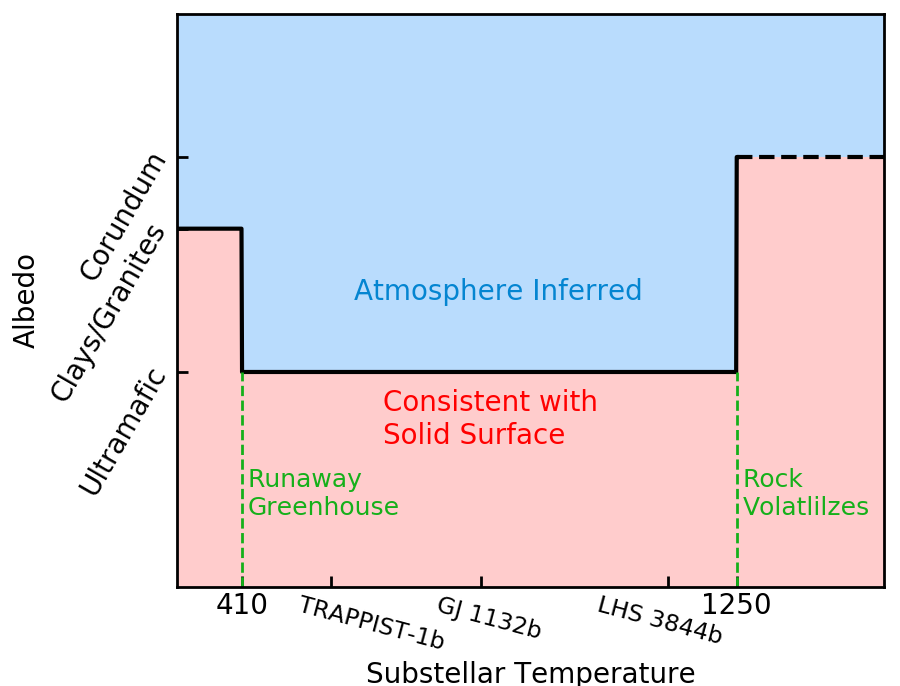}
    \caption{Cartoon demonstrating how measurements of the albedo can determine whether a planet hosts an atmosphere. As we describe in Section~\ref{sec:temprange}, high-albedo, water-rich materials such as clays and granites can form at temperatures below 410~K where planets are not guaranteed to have entered a runaway greenhouse. At temperatures above 1250~K, the rock partially volatilizes. This process may lead to the formation of a high-albedo corundum surface. Between these two extremes, the highest albedo surface that is likely to form is ultramafic, as described in Sections~\ref{sec:albcomp} and \ref{sec:surfaceplausibility}. The blue region indicates where an atmosphere would be inferred. Labels on the x-axis indicate the substellar temperatures of the three planets we consider in detail in this paper, assuming zero albedo and no heat redistribution.}
    \label{fig:cartoon}
\end{figure}

We present another approach for detecting the presence of an atmosphere. For a synchronously rotating rocky exoplanet orbiting a cool host star, observations of the thermal emission constrain the planet's dayside temperature, which can be used to infer its albedo at visible wavelengths by equating the incoming solar radiation to the planet's outgoing radiation. This method has been used previously to infer the albedos of giant exoplanets \citep{CowanAgol2011}. If possible exoplanet surface compositions have relatively low albedos, a high measured albedo would indicate the presence of an atmosphere, as illustrated in Figure~\ref{fig:cartoon}.

This method of atmospheric detection is complementary to the methods described above, because it provides a way to detect thin atmospheres that do not transport enough heat to impact the planet's thermal phase curve or secondary eclipse depth but have high-albedo clouds. Solar system bodies with atmospheres thinner than 1~bar (e.g. Mars) are still able to host significant high-albedo cloud layers with optical depths of order unity, so it is possible that some exoplanet atmospheres will be similar \citep{Smith2008,Clancy2017,Haberle2017}. Additionally, the top of the H$_{2}$SO$_{4}$ cloud deck on Venus is at a pressure level of $\approx1$~bar \citep{Esposito1983}, suggesting that a thin atmosphere with a Venus-like composition could also host high-albedo clouds.

In Section~\ref{sec:temprange} we calculate the range of planet substellar temperatures at which a high-albedo detection points unambiguously to the presence of an atmosphere. We describe our method of calculating planet albedos from thermal infrared observations in Section~\ref{sec:calcalb}. In Section~\ref{sec:results} we discuss the albedos we derive for each of the planet surface compositions we consider, and the characteristics of cloud layers that would be implied by an atmospheric detection with this method. We discuss the surface compositions we expect to exist on these hot rocky planets in Section~\ref{sec:surfaceplausibility}, and list processes that could act to darken or brighten the planet surface in Section~\ref{sec:agentsofdarkness}. We compare our atmospheric detection method to other methods in Section~\ref{sec:familytree} and conclude in Section~\ref{sec:conclude}.

\section{Methods}

\subsection{Planet Substellar Temperature Range}
\label{sec:temprange}

The range of planetary temperatures we consider for this observational technique is limited to zero-albedo substellar temperatures between 410~--~1250~K by two theoretical calculations. We assume no heat redistribution, so the substellar temperature $T_{sub}$ is related to the equilibrium temperature $T_{eq}$ by the equation $T_{eq}=T_{sub}\left(\frac{1}{4}\right)^{1/4}$. The range of substellar temperatures $T_{sub}=410$~--~1250~K corresponds to equilibrium temperatures of 300~--~880~K.

The lower temperature limit is set by the runaway greenhouse limit at zero-age main-sequence luminosity. At temperatures lower than the greenhouse limit the planet's surface could include high-albedo salt flats or water-rich materials such as clays or granites, which would complicate the interpretation of a high-albedo detection. We base our estimate of the runaway greenhouse threshold on the calculations of \citet{Kopparapu2013}. However, \citet{Yang2013} found that clouds on the substellar hemisphere could prevent a planet around an M dwarf from entering a runaway greenhouse state until stellar fluxes twice as large as those reported by \citet{Kopparapu2013}. Therefore, we double the stellar fluxes of the \citet{Kopparapu2013} runaway greenhouse limit to conservatively account for clouds and other factors that may similarly delay the runaway greenhouse \citep{Yang2013,Yang2014,Kodama2018}. For all three planets we consider in this study, this method provides a lower substellar temperature limit of $T_{sub}=410$~K. The conservative runaway greenhouse limit given by this calculation depends on the specific stellar and planetary parameters, but for M dwarfs it will be at stellar fluxes of $\approx2300$\,--\,2500~W/m$^{2}$.

The upper temperature limit is set by the rate at which rock can be partially devolatilized. At high enough temperatures, all components of the rock at the substellar point except corundum will vaporize, leaving behind a high-albedo calcium- and aluminum-rich surface made of materials such as Al$_{2}$O$_{3}$ \citep{Kite2016}. This high-albedo surface would again prevent distinction of a high-albedo atmosphere from a lower-albedo surface, so we limit our study to lower temperatures.

To derive the temperature at which rock devolatilization would impact the overall albedo, we use the MAGMA model of gas-melt chemical equilibrium to calculate the rate at which rock could be devolatilized, assuming a starting composition equal to that of the Earth's continental crust \citep{Fegley1987,Schaefer2009, Kite2016}. Continental crust devolatilizes faster than other possible starting rock compositions because it has a higher vapor pressure, so this choice of starting composition gives a conservative (i.e., low) estimate of the temperature at which devolatilization becomes significant. The MAGMA model outputs the pressure $P$ of the rock vapor over the surface, which we convert to a flux $F$ of rock from the dayside hemisphere to the nightside hemisphere using the equation
\begin{equation}
    F=\frac{c_{s}P}{g},
\end{equation}
where $c_{s}$ is the sound speed and $g$ is the gravitational acceleration. Here we assume that the wind speed of the rock vapor is equal to the sound speed. Models of tenuous vapor atmospheres, both for super-Earth exoplanets and for Jupiter's moon Io, indicate the presence of supersonic winds over a broad region of parameter space, with winds across the terminator typically 2-3 times the sound speed \citep{Ingersoll1985, Castan2011}.

We then convert this flux of material over the terminator to a rate $R$ of vaporization from the dayside hemisphere using the equation
\begin{equation}
    R=\left(\frac{F}{\rho}\right)\left(\frac{2\pi R_{p}}{4 \pi R_{p}^{2}}\right)=\frac{F}{2R_{p}\rho},
\end{equation}
where $R_{p}$ is the planet radius and $\rho$ is the density of the rock. 

\begin{figure}
    \centering
    \includegraphics[width=\linewidth]{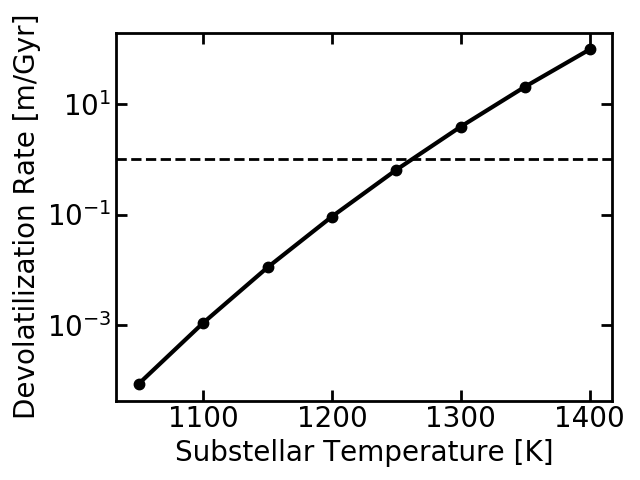}
    \caption{Devolatilization rate as a function of temperature.  The horizontal dashed line shows an order-of-magnitude estimate of the inner solar system rate of meteoritic gardening, which stirs fresh material to the surface. Values below this line indicate temperatures at which the rock composition is little-affected by devolatilization. The devolatilization rate is calculated using MAGMA \citep{Fegley1987,Schaefer2009}, for continental crust composition. Other plausible rocky planet crust compositions would have even lower devolatilization rates.
    }
    \label{fig:devol}
\end{figure}

Figure~\ref{fig:devol} shows the rate of devolatilization as a function of temperature. We compare this rate to the rate of meteoritic gardening, which determines how quickly fresh, low-albedo material could be mixed from below the rock surface \citep{Melosh2011}. We assume that impact gardening mixes regolith to a depth of order $1$~m/Gyr based on analogy to the solar system \citep{Warner2017, Fassett2017}, but our calculations are insensitive to the exact impact gardening rate to within an order of magnitude because the rock vapor pressure increases very rapidly with increasing substellar temperature (Figure~\ref{fig:devol}). The rate of devolatilization will be slower than meteoritic gardening for substellar temperatures $\leq1250$~K. We find that this is approximately the temperature at the Roche lobe radius for an Earth-density planet orbiting a mid M dwarf, so all close-in planets around M dwarfs will be cool enough to avoid partial devolatilization.

\subsection{Observed Planetary Fluxes and Albedos}

\label{sec:calcalb}

We investigate a variety of potential rock compositions with different albedo properties. We consider the eight compositions outlined in \citet{Hu2012}: basaltic, clay, feldspathic, Fe-oxidized (50\% nanophase hematite, 50\% basalt), granitoid, ice-rich (50\% ice and 50\% basalt), metal-rich (FeS$_{2}$), and ultramafic. \citet{Hu2012} created reflectance spectra for three of these surfaces using laboratory measurements of rock powders, and for the other five surfaces used radiative-transfer modeling based on laboratory samples of component minerals. These model spectra assume relatively fine-grained rocks. Fine grains generally have higher albedos than coarse grains, so the albedos taken from \citet{Hu2012} represent conservative upper limits of the albedos of these eight compositions.

\begin{figure*}
    \centering
	\includegraphics[width=\linewidth]{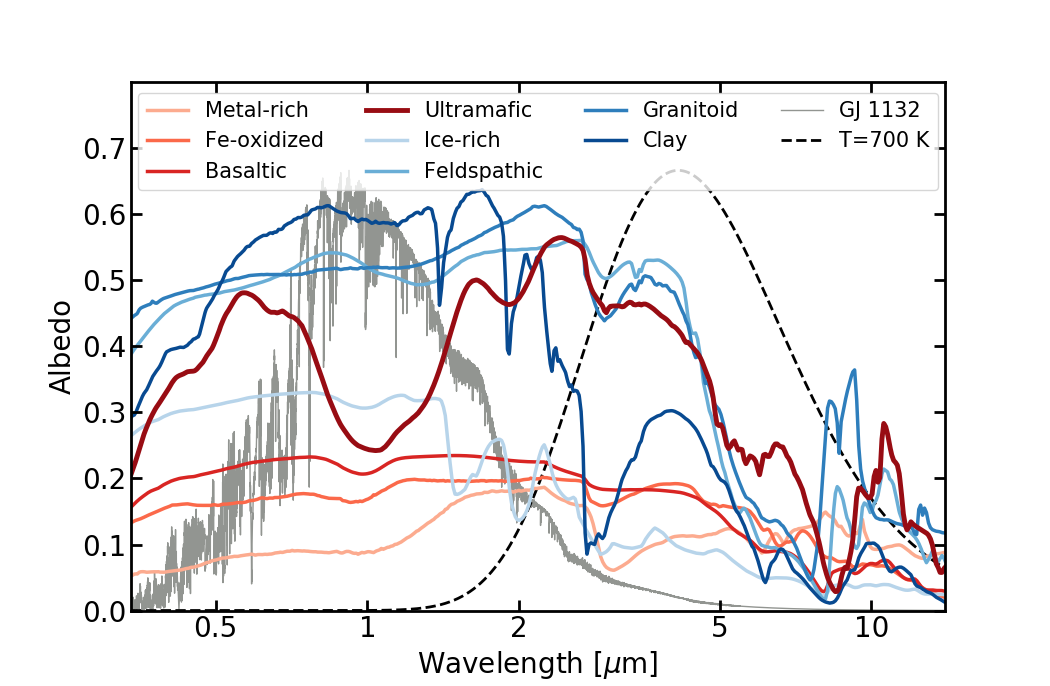}
    \caption{Albedo as a function of wavelength for the eight types of planetary surfaces we consider in detail in this paper, taken from \citet{Hu2012}. The solid grey line shows a PHOENIX model for the stellar spectrum of GJ~1132 \citep{Husser2013}, and the dashed black line shows a blackbody at $T=700$~K, which is the approximate temperature of the dayside of GJ~1132b. Red tinted lines indicate surface compositions that are more plausible for planets in $T=410$-1250~K orbits, while blue tinted lines are compositions that are not likely to occur at these temperatures. The thick, dark red line indicates the reflectance spectrum of ultramafic rock, which is discussed in more detail throughout the paper because it has the highest albedo of the plausible surfaces, and so it the limiting case for atmosphere identification using the method proposed in this paper.}
    \label{fig:albedos}
\end{figure*}

Figure \ref{fig:albedos} shows the albedos of these surfaces as a function of wavelength \citep{Hu2012}. The fundamental basis for the overall shapes of these spectra is that many rock-forming minerals have strong spectral slopes in the $0.4$-$5$~$\mu$m range. We discuss the plausibility of these surfaces forming on terrestrial planets with $T_{sub}=410$-$1250$~K in Section \ref{sec:surfaceplausibility}.

For each rock composition, we determine the temperature of a hypothetical planet with a surface of that composition by setting its outgoing flux equal to the absorbed flux it receives from its star. The absorbed flux from the star as a function of wavelength, $F_{\star}(\lambda)$, is given by
\begin{equation}
F_{\star}(\lambda)=\left[F_{\lambda}(1-\alpha_{\lambda})\right]\left(\frac{R_{\star}^{2}}{a^{2}}\right),
\end{equation}
where $R_{p}$ is the planetary radius, $R_{\star}$ is the stellar radius, $a$ is the distance from the planet to the star, $\alpha_{\lambda}$ is the planet's albedo as a function of wavelength, and $F_{\lambda}$ is the spectral flux density in W~m$^{-3}$ from a PHOENIX model for the star \citep{Husser2013}. The flux emitted by the planet is approximated by
\begin{equation}
    F_{p}(\lambda)=\pi B_{\lambda}(T_{day})(1-\alpha_{\lambda}),
\end{equation}
where $B_{\lambda}(T)$ is the flux emitted by a blackbody and $T_{day}$ is the planet dayside brightness temperature, which is related to the substellar temperature $T_{sub}$ through the equation $T_{day}=\left(\frac{2}{3}\right)^{1/4}T_{sub}$. The factor of $\frac{2}{3}$ assumes zero heat redistribution and a bare rock surface \citep{Hansen2008}.\footnote{Even a magma ocean would not redistribute much heat if heated only by the star, and confined to the dayside \citep{Kite2016}.} We also assume here that the planet is in 1:1 spin:orbit resonance, because we are considering hot worlds in close-in orbits around K and M dwarfs. Additionally, approximating the entire dayside as a single blackbody implicitly assumes that the dayside surface is completely covered by one rock type. We integrate these two equations over wavelength and iterate until they are equal to determine the planetary temperature.

For each of these planet surfaces, we sum the light reflected and emitted by the planet to get a planet spectrum as a function of wavelength. We use PandExo to simulate observations of these planets with \textit{JWST} \citep{Batalha2017}. We simulate sets of five secondary eclipse observations using the Mid-Infrared Instrument's Low-Resolution Spectroscopy (MIRI LRS) slitless mode to observe between 5 and 12 $\mu$m. We integrate over this entire wavelength range to produce one broadband planetary flux measurement.

We calculate the planetary brightness temperature that would be inferred from these observations by inverting the equation
\begin{equation}
    \frac{F_{p}}{F_{\star}}=\left(\frac{R_{p}}{R_{\star}}\right)^{2}\int_{\lambda=5\text{ $\mu$m}}^{12\text{ $\mu$m}}\left(\frac{ B_{\lambda}(T_{day,obs})}{B_{\lambda}(T_{\star})}\right)\,d\lambda,
    \label{eq:tinf}
\end{equation}
where $\frac{F_{p}}{F_{\star}}$ is the observed broadband planet-to-star flux ratio and $T_{day,obs}$ is the planet dayside brightness temperature inferred from the observations. Note that this equation assumes that the planet's emissivity ($\epsilon_{\lambda}=1-\alpha_{\lambda}$) is unity. We make this assumption when interpreting our observed planet flux because the planet's emissivity cannot be known a priori. We approximate the star's flux as a blackbody for this calculation because the PHOENIX model spectra only extend to a wavelength of 5~$\mu$m. We then convert this temperature into an inferred planetary albedo using the equation
\begin{equation}
    T_{day,obs}=T_{\star}\sqrt{\frac{R_{\star}}{a}}\left[\frac{2}{3}(1-\alpha_{obs})\right]^{1/4},
    \label{eq:alb}
\end{equation}
where $\alpha_{obs}$ is the albedo inferred from the observations. This equation again assumes unit emissivity.

We calculate the inferred albedo for each of the eight planet surfaces for three planets: the canonical high signal-to-noise planet GJ~1132b \citep{Berta2015}; TRAPPIST-1~b, which orbits a very small star and so is a relatively high-signal transiting planet with an equilibrium temperature near the lower end of our temperature range \citep{Gillon2017}; and the newly discovered \textit{TESS} planet LHS~3844b, which is representative of the type of planets the \textit{TESS} mission will continue to discover \citep{Vanderspek2018}. These three planets together span almost the entire temperature range from 410-1250~K. Table \ref{tab:parameters} lists the details of each planet we consider.

\begin{deluxetable}{ll}
\tablecaption{Stellar and planetary parameters for the three systems we consider.}
\tablecolumns{2}
\tablehead{
\colhead{Parameter} &
\colhead{Value}
}
\startdata
\textit{TRAPPIST-1b} & \\
Star radius & 0.121 $R_\sun$ \\
Star effective temperature & 2511 K\\
Star K magnitude & 10.296 \\
Planet radius & 1.12 $R_\Earth$ \\
Planet orbital period & 1.51 d \\
Planet dayside temperature\tablenotemark{a} & 508 K \\
\textit{GJ\,1132b} & \\
Star radius & 0.207 $R_\sun$ \\
Star effective temperature & 3270 K\\
Star K magnitude & 8.322 \\
Planet radius & 1.16 $R_\Earth$ \\
Planet orbital period & 1.63 d \\
Planet dayside  temperature\tablenotemark{a} & 737 K \\
\textit{LHS\,3844b} & \\
Star radius & 0.189 $R_\sun$ \\
Star effective temperature & 3036 K\\
Star K magnitude & 9.145 \\
Planet radius & 1.32 $R_\Earth$ \\
Planet orbital period & 0.46 d \\
Planet dayside temperature\tablenotemark{a} & 1024 K \\
\enddata
\label{tab:parameters}
\tablenotetext{a}{Assumes no heat redistribution and zero albedo.}
\end{deluxetable}

\section{Results}
\label{sec:results}

We find that all plausible surface compositions for planets in $T_{sub}=410-1250$~K orbits have low albedos, and that even very thin atmospheres can host enough clouds to raise the albedo above that for a bare surface. Our main results depend on the relationship between the actual planet Bond albedo at visible wavelengths and the albedo inferred from observations at mid-infrared wavelengths, which we describe in detail in Section~\ref{sec:albcomp}. We find that the albedo inferred from such observations is lower than the actual Bond albedo for all the surfaces we consider. In Section~\ref{sec:itsamie} we calculate the properties of clouds that have high enough albedos to be distinguishable from bare rock surfaces.

\subsection{Comparison of Inferred and Actual Planetary Albedos}
\label{sec:albcomp}

A key complication in relating an inferred albedo to the presence of an atmosphere is the difference between the inferred planet albedo and its actual Bond albedo. A realistic surface with a non-constant albedo at the wavelengths where it emits radiation will have an albedo inferred from observations of planetary radiation that differs from its true Bond albedo (the actual percentage of starlight at shorter wavelengths that is reflected off the planet's surface). This is related to Kirchhoff's law of thermal radiation: a planet with an albedo that changes as a function of wavelength will emit relatively more or less light at certain wavelengths compared to a blackbody with a constant emissivity at all wavelengths.

\begin{figure}
	\begin{subfigure}{}
	\includegraphics[width=\linewidth]{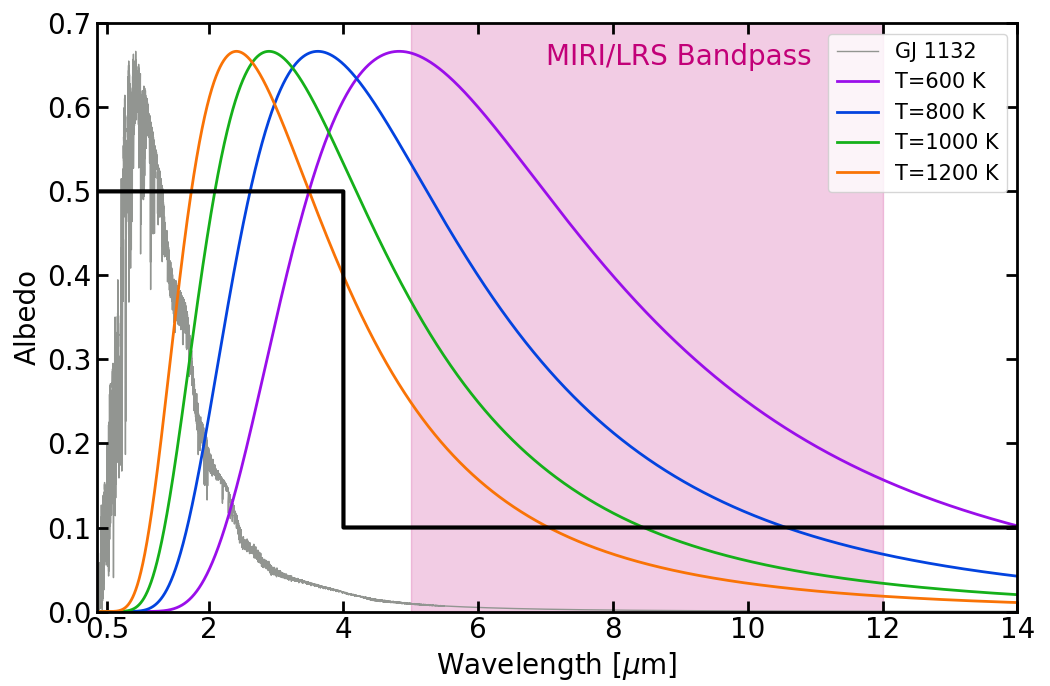}
	\end{subfigure}
	\vspace{-3mm}
	\begin{subfigure}{}
	\includegraphics[width=\linewidth]{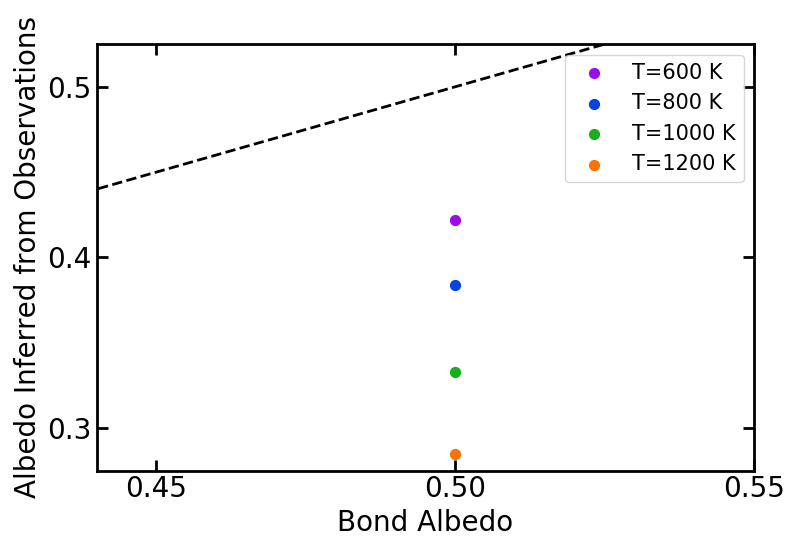}
	\end{subfigure}
\caption{\label{fig:diffplanettemps} Simplified example showing how the inferred albedo can be different from the Bond albedo for a planet where the albedo changes as a function of wavelength. The upper plot shows the simple step function albedo (black line) overplotting the stellar spectrum for GJ~1132 and example planet blackbodies at four different temperatures ranging from $T=600$~K to $T=1200$ K. The magenta shaded region indicates the MIRI/LRS bandpass. The lower plot shows the albedo inferred from \textit{JWST}/MIRI observations of each planet. The dashed line indicates where the inferred albedo equals the Bond albedo.}
\end{figure}

Figure~\ref{fig:diffplanettemps} demonstrates why inferred albedo differs from Bond albedo for a simple example where the albedo is a step function given by
\begin{equation}
    \alpha_{\lambda}= 
    \begin{cases}
        0.5, & \lambda<4 \mu \text{m} \\
        0.1, & \lambda>4 \mu \text{m} \\
    \end{cases}.
\end{equation}
The lower panel of this figure shows the actual Bond albedo compared to the albedo inferred from observations with \textit{JWST}/MIRI for a set of planets at different temperatures spanning the range we consider. For all of the planets, the step function means that the MIRI bandpass (5-12~$\mu$m) is at a lower albedo (and thus higher emissivity) than shorter wavelengths. Therefore, in order to satisfy energy balance, the planet must emit relatively more of its flux at the long wavelengths where the emissivity is higher than if it were emitting as a blackbody with a constant emissivity. This means the planet will appear to be at a higher temperature in the MIRI bandpass, and so the inferred albedo will be lower than the actual Bond albedo. For planets at higher temperatures, this effect is even stronger because more of the planet's emission is at low-emissivity short wavelengths. As a result, in order to satisfy energy balance, the relative amount emitted at longer wavelengths is even higher compared to a constant-emissivity blackbody.

\begin{figure}
\centering
    \begin{subfigure}{}
	\includegraphics[width=0.7\linewidth]{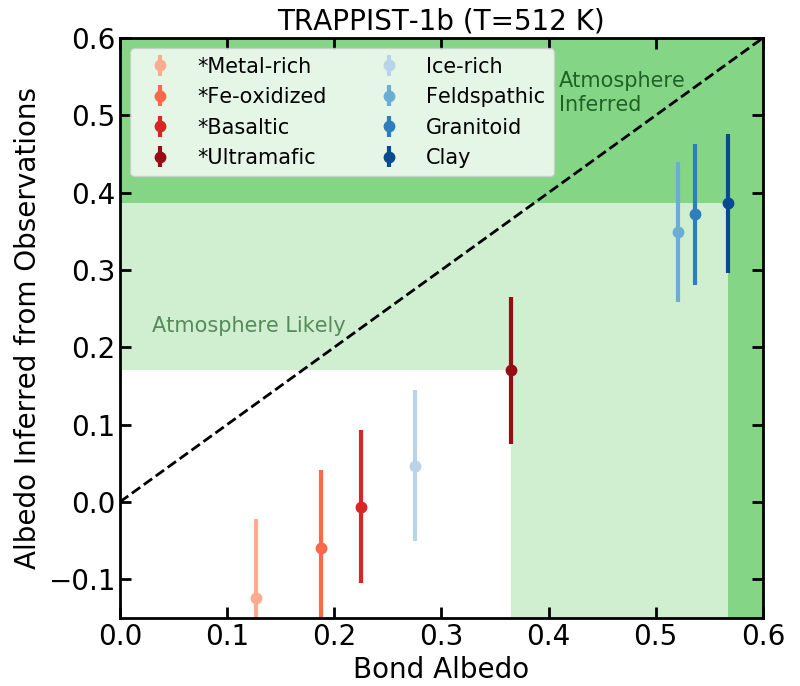}
	\end{subfigure}
	\begin{subfigure}{}
	\includegraphics[width=0.7\linewidth]{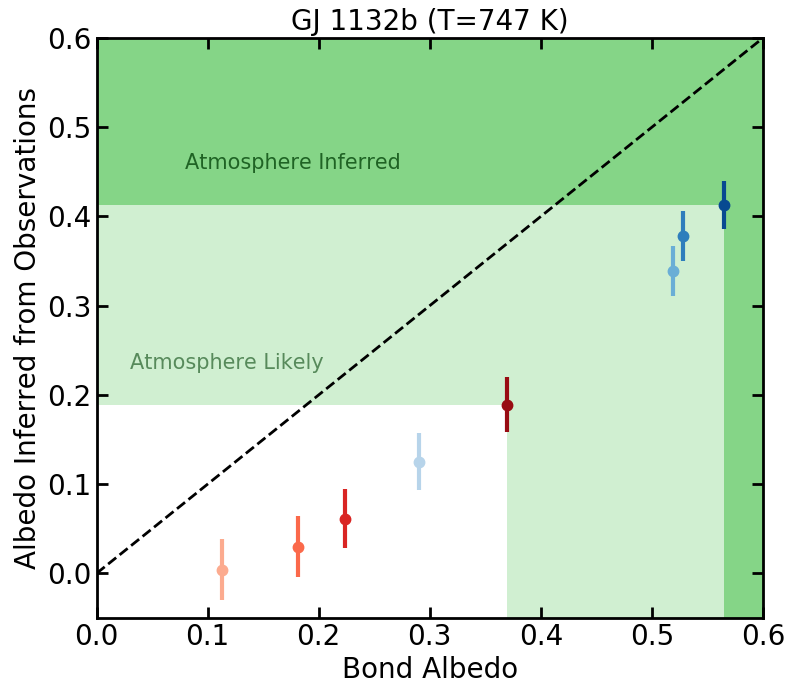}
	\end{subfigure}
	\begin{subfigure}{}
	\includegraphics[width=0.7\linewidth]{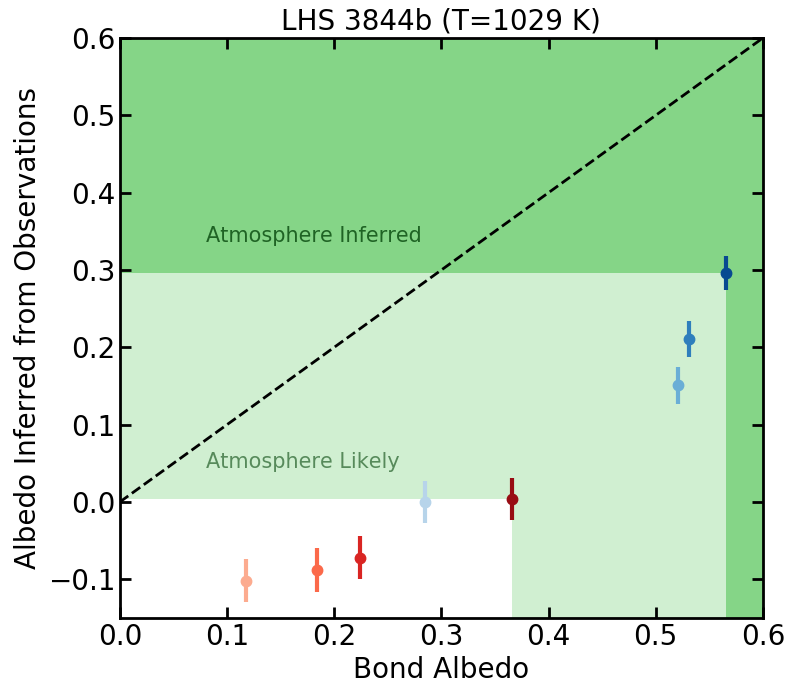}
	\end{subfigure}
\caption{\label{fig:obsvsactual}Insolation flux-weighted albedo of the eight possible planetary surfaces in the wavelength range from $0.1$-$3.5$~$\mu$m compared to the albedo inferred from longer-wavelength observations with \textit{JWST}/MIRI for TRAPPIST-1~b (upper panel), GJ~1132b (middle panel), and LHS~3844b (lower panel). Temperatures quoted in the plot titles assume $\alpha=0$ and no heat redistribution. The error bars indicate $1\sigma$ observational uncertainty for five stacked secondary eclipse observations. The black dashed line shows where the Bond albedo equals the inferred albedo. In all cases, the inferred albedo is lower than the actual albedo. Note that in some cases the inferred albedo appears to be negative. This is due to the assumption of unit emissivity when calculating the albedo. The light and dark green shaded regions indicate where the albedo is high enough that an atmosphere is likely and where one is needed to explain the observation, respectively.}
\end{figure}

Figure \ref{fig:obsvsactual} shows a comparison of the insolation flux-weighted albedos of the eight planetary surfaces at shorter wavelengths ($0.1$-$3.5$~$\mu$m, the range in which all three of the M dwarf stars we consider emit $>90$\% of their flux) to their inferred albedos from broadband mid-infrared observations with \textit{JWST}/MIRI. The error bars represent $1\sigma$ observational uncertainties for a set of five stacked secondary eclipses. In all cases the inferred albedo is lower than the actual Bond albedo because the spectra generally have higher albedos at shorter wavelengths and lower albedos at longer wavelengths.

The surfaces can be grouped into two rough categories based on the shapes of their spectra (Figure \ref{fig:albedos}). First, feldspathic, granitoid, ultramafic, and clay generally show a high albedo at short wavelengths, then an abrupt transition around 2-5 $\mu$m to low albedo at longer wavelengths (Figure \ref{fig:albedos}). All of these surfaces also have a lower emissivity at 3-5 $\mu$m (the peak of the flux for a planet at the temperature of GJ~1132b if that planet emitted as a blackbody) than at 5-12 $\mu$m (the MIRI bandpass). This means they will emit a larger percentage of their flux in the MIRI bandpass, leading to a higher inferred temperature and a lower inferred albedo. The larger the increase from the 3-5 $\mu$m emissivity to the 5-12 $\mu$m emissivity, the lower the inferred albedo will be relative to the Bond albedo. Feldspathic surfaces have the largest difference in its emissivity in these two wavelength ranges, followed by clay, then granitoid, then ultramafic. Therefore, in this set of four surfaces, the feldspathic surface shows the largest deviation from the 1:1 line in Figure \ref{fig:obsvsactual}, and the ultramafic shows the smallest deviation.

The second category consists of ice-rich, basaltic, Fe-oxidized, and metal-rich surfaces. All of these surfaces can be approximated as a more gradual slope to lower albedos at longer wavelengths, without the sharp transition of the first four surfaces (Figure \ref{fig:albedos}). Ice-rich is on the edge between the two categories, but its sharp drop-off is smaller and at shorter wavelengths. The metal-rich surface has an albedo that is close to constant, so its inferred albedo should be close to its Bond albedo and it should fall closest to the 1:1 line in Figure \ref{fig:obsvsactual}. The other three surfaces all have slightly higher emissivity at the MIRI wavelengths, so they should again all be slightly farther from the 1:1 line. Among those three surfaces, the Fe-oxidized surface has the smallest difference between its visible and MIRI emissivity, followed by ice-rich, then basaltic. So in this group of three surfaces, Fe-oxidized is closest to the 1:1 line and basaltic is farthest.

The stars GJ~1132 and LHS~3844 have nearly the same PHOENIX spectra because their effective temperatures only differ by 200~K. The Bond albedos for these two planets differ by $<0.01$. Therefore, the primary difference between GJ~1132b and LHS~3844b is that LHS~3844b is much warmer than GJ~1132b (zero-albedo, zero-redistribution $T_{day}=$\,1024~K for LHS~3844b as opposed to 737~K for GJ~1132b), so that LHS~3844b emits its flux at slightly shorter wavelengths. The overall effect is that a smaller percentage of the planet’s blackbody curve is in the high emissivity regions at longer wavelengths, so the warmer planet will emit relatively more at these wavelengths and the inferred albedos will appear even lower than for the case of GJ~1132b. For all of the surfaces except clay, the surface albedos at the peak of GJ~1132b’s blackbody flux and that of LHS~3844b are within 0.04 of each other, so those surfaces all uniformly have lower inferred albedos. For the clay surface, the peak flux of LHS~3844b happens to be emitting in a region where the albedo is almost 0.2 lower than the surrounding parts of the spectrum. This means the peak flux for the clay surface is at a higher emissivity, so relatively less flux needs to be emitted at the MIRI wavelengths. Therefore the clay surface has a higher inferred albedo relative to the other surfaces compared to what it had for GJ~1132b.

The star TRAPPIST-1 is 700~K cooler than GJ~1132, but the difference in inferred albedos for TRAPPIST-1~b and GJ~1132b is again primarily due to the different planet temperatures (and not due to the difference in stellar effective temperature). TRAPPIST-1~b has a temperature of $\approx470$~K, which is cool enough that the peak of its flux is emitted in the MIRI bandpass. Therefore, the difference between the inferred albedo and the true Bond albedo simply depends on the difference between the emissivity at short wavelengths, where the starlight is absorbed, and at long wavelengths, where the planet emits its flux. A surface with a larger difference between its emissivity at wavelengths $<3.5$~$\mu$m and its emissivity at $5-12$~$\mu$m will have a larger difference between its inferred and Bond albedos. Within the first category of surfaces, feldspathic has the largest difference between its short-wavelength and long-wavelength emissivities, followed by granitoid, ultramafic, and clay. Therefore, among these four surfaces, feldspathic has the largest difference between its inferred and Bond albedos. Similarly, the basaltic and metal-rich surfaces have the largest and smallest difference in emissivities among the second category of surfaces, so they have the largest and smallest difference between inferred and Bond albedos, respectively. 

Our method of differentiating a bare rock surface from an atmosphere relies on the fact that bare surfaces have generally low albedos, so any high-albedo detection would have to come from an atmosphere. Our calculations indicate that observations of a bare rock surface would lead to inferring a lower albedo than that of the real surface. Additionally, despite the offset between Bond albedo and inferred albedo, low Bond albedo surfaces always lead to lower inferred albedos, and high Bond albedo surfaces always lead to higher inferred albedos. This means that the detection of a low secondary eclipse depth corresponding to a high inferred albedo above about 0.4 would unambiguously indicate an atmosphere. As discussed in Section~\ref{sec:agentsofdarkness}, this conclusion is robust to considering processes that could darken or brighten the surface, which were not considered in \citet{Hu2012} 

\subsection{Distinguishing Planetary Surfaces from High-Albedo Clouds}
\label{sec:itsamie}

Given that an inferred high albedo can indicate the presence of an atmosphere, what can be said about that atmosphere? Although Rayleigh scattering alone can in principle cause an albedo $>0.5$, in practice a more likely cause of high albedo is clouds. For example, Venus's high albedo of 0.7 is due to clouds.

We calculated the albedos of clouds with a variety of properties to determine what types of atmospheres could be distinguished from bare rock surfaces on the basis of albedo alone. We constructed a grid of atmospheres with cloud column masses between $10^{-10}-10^{1}$~g/cm$^{2}$ and non-absorbing cloud particles with radii between $10^{-1}-10^{1.7}$~$\mu$m. From these parameters we calculated the optical depth ($\tau$) using the equation
\begin{equation}
    \tau=\frac{3 Q m_{col}}{4\rho_{p}r_{p}},
\end{equation}
where $Q$ is the scattering efficiency, $m_{col}$ is the cloud column mass, and $\rho_{p}$ and $r_{p}$ are the particle density and radius, respectively \citep{Pierrehumbert2010}. We then calculated the albedo ($\alpha$) using the equation
\begin{equation}
    \alpha=\alpha_{a}+\frac{(1-\alpha_{a}^{'})(1-\alpha_{a})\alpha_{g}}{1-\alpha_{a}^{'}\alpha_{g}},
\end{equation}
where
\begin{equation}
    \alpha_{a}^{'}=\frac{(1-g)\tau}{1+(1-g)\tau},
\end{equation}
\vspace{0.5 mm}
\begin{equation}
    \alpha_{a}=\frac{\frac{-1}{2}\beta+(1-\hat{g})\tau}{1+(1-\hat{g})\tau},
\end{equation}
and $\beta=1-\mathrm{e}^{-\tau}$ \citep{Pierrehumbert2010}. In these equations, $\hat{g}$ is the asymmetry factor and $\alpha_{g}$ is the albedo of the rock surface. \citet{Mbarek2016} calculated equilibrium chemistry cloud compositions for secondary atmospheres at $T=410$-$1250$~K, and several of these possible compositions (including K$_{2}$SO$_{4}$, KCl, and Na$_{2}$SO$_{4}$) have indices of refraction $n_{R} \approx 1.5$ and $n_{I} \approx 0$ \citep{Querry1987,CRC2005}. Therefore, we used values of $Q$ and $\hat{g}$ for particles with indices of refraction $n_{R}=1.5$ and $n_{I} \approx 0$ (values range from $0.2<Q<4$ and $0.2<\hat{g}<0.8$).

\begin{figure}
    \centering
    \includegraphics[width=\linewidth]{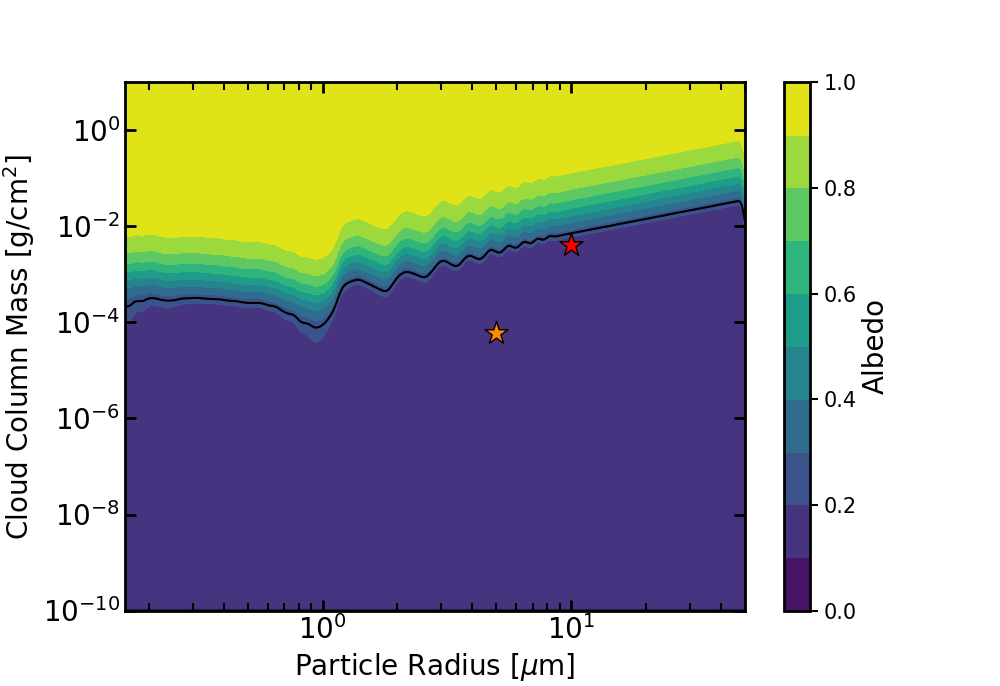}
    \caption{Contour plot of total planet albedo as a function of cloud particle radius and cloud column mass. The area above the black line is the region of parameter space where the albedo is more than $2\sigma$ greater than the inferred bare rock albedo of an ultramafic rock surface on GJ~1132b. The red and orange stars indicate typical cloud parameters for Earth \citep{Wallace2006} and Mars \citep{Clancy2017}, respectively.}
    \label{fig:cloud}
\end{figure}

Figure \ref{fig:cloud} shows a contour plot of potential albedos for GJ~1132b, assuming an ultramafic rock surface. The area above the black line is the region of parameter space where the total planet albedo is more than $2\sigma$ higher than the inferred albedo of the bare rock surface. A detection of an albedo higher than this value would suggest the presence of an atmosphere on this planet. For this surface, an atmosphere with a cloud column mass greater than $8 \times 10^{-5}$--$3 \times 10^{-2}$~g/cm$^{2}$ ($\tau>0.8-5$) would have a higher albedo. The stars on Figure~\ref{fig:cloud} indicate typical cloud parameters for Earth \citep{Wallace2006} and Mars \citep{Clancy2017}. While clouds made of larger, Earth-sized particles would be harder to detect using our method, smaller-particle hazes would be detectable at lower cloud column masses.

Most of the surface types exhibit a similar behavior, with column masses greater than $4 \times 10^{-5}$ -- $5 \times 10^{-2}$~g/cm$^{2}$ ($\tau>0.5$ -- $7$) having high enough albedos to suggest an atmosphere. The minimum pressure required to support such high-albedo cloud layers is significantly smaller than the $\approx1$~bar required to transport heat. For example, Mars has regionally extensive high-albedo clouds made of both H$_{2}$O and CO$_{2}$ ice in a 6~mbar atmosphere \citep{Smith2008,Haberle2017}. These clouds can be optically thick, especially near surface ice deposits.

Our calculation implicitly assumes that a large portion of the dayside is covered in clouds so that the disk-integrated dayside albedo is large. This is a reasonable assumption for the close-in terrestrial exoplanets we consider, because 3D global climate simulations of synchronously rotating planets exhibit upwelling and cloud cover over much of the dayside, with downwelling and clear skies confined to the nightside \citep{Yang2013}.

\section{Discussion}

\subsection{Which Surface Compositions Are Expected to Exist?}
\label{sec:surfaceplausibility}

A high surface albedo can produce a low secondary eclipse depth, so using the secondary eclipse depth to screen for the presence of atmospheres will only work if there is a prior constraint on the distribution of possible surface albedos. We have two primary sources of information on the albedos of terrestrial planets: observations of rocky objects in the solar system and laboratory spectra of geologically plausible surfaces.

\subsubsection{Surfaces Observed in the Solar System}

Table~\ref{tab:ssalb} lists the albedos of several solar system bodies. The rocky bodies in the solar system generally have low albedos \citep{MaddenKaltenegger2018}. E-type asteroids\footnote{E-types are common between Mars and the inner edge of the main asteroid belt (Hungaria region). E-types are plausibly leftovers from the formation of Earth and Mars.} are an interesting exception to the general trend of dark solar system rocks, with albedos $>0.3$. One of the brightest E-type asteroids, 44 Nysa, is listed in Table~\ref{tab:ssalb} for comparison \citep{Takahashi2011}. E-types are likely the source of enstatite chondrite meteorites. If this mapping between meteorite type and asteroid type is correct, then the cause of the high asteroidal albedo is that the rocks record very reducing conditions - so any iron gets reduced to Fe, and there is very little Fe$^{2+}$ in the silicate (Fe and other transition metals being a big cause of the dark color of the most common silicate rocks). This matters because the reducing conditions are analogous to those expected for evaporated cores (which will be discussed further in Section~\ref{sec:jedi}) if enough hydrogen was originally present to overwhelm buffering by  Fe-oxides. Although it is unclear what an evaporated core will look like because we have never imaged one, this redox similarity leads us to speculate that E-types may be the best solar system analog to the surface composition of evaporated cores.

\begin{table}
    \centering
    \begin{tabular}{c|c}
        Body & Bond Albedo  \\
        \hline
        Mercury & 0.07 \\
        Moon & 0.11 \\
        Mars & 0.25 \\
        4 Vesta & 0.18 \\
        1 Ceres & 0.03 \\
        Io & 0.6 \\
        44 Nysa & 0.33 \\
    \end{tabular}
    \caption{Bond albedos of solar system bodies \citep{MaddenKaltenegger2018,Takahashi2011}.}
    \label{tab:ssalb}
\end{table}

The near-infrared reflectance spectrum of enstatite is shown in Figure~\ref{fig:jedisurfaces}. At the wavelengths where M dwarfs emit most of their light, enstatite has an albedo between 0.3-0.4 that is similar in shape and magnitude to that of ultramafic rock. Therefore, observations of an enstatite surface would likely lead to a similar inferred albedo as that of an ultramafic surface. Our method of calculating inferred albedo would still allow detection of a high-albedo atmosphere on a planet with an enstatite surface.

Although sulfur species (mainly SO$_{2}$) are responsible for the high albedo of Io, SO$_{2}$ would not be condensed as a solid on the surface of the hot planets we consider here and sulfur would be liquid (for $T>$ 427K) and quite dark \citep{Nelson1983}.

The solar system contains only one world in a $T>410$~K orbit -- the $\sim$0.06 $M_{\oplus}$ world Mercury.  In principle exoplanet albedo measurements could be used to supplement solar system data, to build up an empirical prior on rocky planet surface albedo. However, so far direct measurements of rocky-exoplanet albedos are limited \citep{Rouan2011,SheetsDeming2014,SheetsDeming2017,JansenKipping2018}. Moreover, even to use the solar system data for exoplanets in hotter orbits we need to think about how the albedo would (or would not) be affected by processes at work on a hotter orbit. Therefore, our primary focus in this work is on laboratory spectra of hypothetical planet surface compositions, as well as processes that would make those compositions more or less likely.

\subsubsection{Laboratory Spectra of Hypothetical Surfaces}

We calculate albedos for eight possible surfaces in this paper, but not all of these surfaces are likely to form at the high temperatures we consider. Many planetary surface types require water to form. For example, it is roughly true that ``[n]o water, no granites - no oceans, no continents'' \citep{CampbellTaylor1983}. Forming granites on Earth involves water. Water is difficult to accrete, and difficult to retain, in a $T_{sub}=$~410-1250~K orbit. As long as water is abundant at the surface, a runaway greenhouse climate is expected, and this will favor H escape to space \citep{Hamano2013}. If the water is retained somehow, then the planet will have a H$_2$O vapor atmosphere. 

Several other high-albedo surface types also require water to form \citep{McSween2019}. Clays need water to form, either as structural water or for the leaching weathering reactions that produce anhydrous phyllosilicates such as kaolinite. Salt flats such as those found in White Sands, New Mexico and Salar de Ayuni, Bolivia also require water to form. Although pure quartz sand (SiO$_2$) on Earth can be found in deserts, it again is a signature of water - desert sand is a breakdown product of high-Si crust, often weathered and physically concentrated by processes involving liquid water. Small amounts of high-Si rock can be made without water by partial remelting of basalt, but this is unlikely to cover the entire planetary surface. 

Feldspathic (plagioclase-feldspar-dominated) flotation crust can form on dry worlds. While feldspathic flotation crusts are a possibility for dry Moon-sized worlds, they are less likely for larger exoplanets \citep{ElkinsTanton2012}. The only feldspathic crust we know is the Lunar highlands. The standard story of origin for the Lunar highlands crust involves formation of plagioclase as a liquidus phase in a cooling magma ocean \citep{Elkins2011}. However, plagioclase will not crystallize from a mafic or ultramafic melt at pressure much above 1 GPa, so for a large Earth-sized planet it will only form during the last dregs of magma-ocean crystallization. Moreover the only feldspathic flotation crust we know of, the Moon, is not as high albedo as the plagioclase-feldspar laboratory spectra might suggest because of processes such as space weathering (see Section \ref{sec:sith}).

Several of the other planetary surface types we consider, including basalt and ultramafic rock, can form without water. Planetary crusts are primarily composed of basalt, a dark rock type with 45-53 wt\% SiO$_2$. Basalt is very common in the solar system because it is the expected product of low-percentage ($\sim$10\%) partial melting of ``average rock''\footnote{We assume that solar system mantle rock compositions -- which upon melting, yield basalt -- are representative of rock elsewhere in the Universe. White dwarf data are consistent with Earth-like Mg/Si ratios \citep{JuraYoung2014} and stellar Mg/Si ratios show little scatter in the solar neighborhood \citep{Bedell2018}. However, we are not aware of any work to study how Mg/Si variability propagates into the percentage of partial melting nor the mineralogy of the resulting lavas. Therefore, this assumption is unverified, but could be modeled in the future using existing datasets \citep{UnterbornPanero2017,HinkelUnterborn2018}.} \citep{BVTP1981,TaylorMcLennan2009}. Basalt has a relatively low albedo.

Relative to basalt, ultramafic rocks have a higher albedo and are a potential false positive for atmosphere detection using the secondary-eclipse-depth technique in all orbits, including those with $T={sub}$~410-1250~K. Ultramafic rocks (which have $<$45 wt\% SiO$_2$) are the result of high-percentage \citep[$>$30\%;][]{GroveParman2004} partial melting of ``average rock''. Such high-percentage partial melts are expected for worlds with high mantle temperatures, including strongly tidally heated worlds and young worlds with strong radiogenic heating \citep{Kite2009}. Geologic terrains from the first 2 Gyr of Earth history often contain ultramafic rocks (specifically komatiites) because they correspond to a time when Earth's mantle was hotter than it is today, and so partial melt fractions were higher \citep{Herzberg2010}. Although ultramafic rocks can be dark in outcrop, the dominant minerals in ultramafic rocks -- pyroxene and olivine -- have high reflectance in the visible \citep{Hu2012}. The possibility of ultramafic surfaces on Earth-sized exoplanets is the limiting case for using albedo to detect atmospheres on worlds in $T={sub}$~410-1250~K orbits, as illustrated in Figure~\ref{fig:cartoon}. It is hard to make more reflective surfaces for these worlds, so higher albedos imply an atmosphere. Other surface types shown in Figure \ref{fig:albedos}, including Fe-oxidized, basaltic, and metal-rich, have lower albedos, and so they are not worrisome for the purposes of screening for atmospheres using secondary eclipse depth.

\subsection{Factors That Could Affect Surface Albedo}
\label{sec:agentsofdarkness}

There are several processes that could act to make the observed surfaces darker or brighter. We discuss these possibilities below.

\subsubsection{Darkening Processes}
\label{sec:sith}

Solar system worlds can be darkened by minor contaminants, which are not considered in the spectra shown in Figure \ref{fig:albedos}. For example, Mercury's surface is very dark, likely due to minor graphite \citep{Izenberg2014}. E-type asteroids are also likely darkened by minor contaminants. Even though they are among the highest albedo rocky objects in the solar system ($\alpha \geq 0.3$), the bulk mineralogy of enstatite chondrites suggests that E-type asteroids should have even higher albedos. Grain size and texture effects can also impact how much a surface is darkened \citep{Carli2015}. Darkening effects would strengthen the conclusion that a high-albedo detection is due to an atmosphere, because the surface would be expected to be even darker.

Space weathering also darkens surfaces \citep{Brunetto2015, Domingue2014}. The space weathering effect on bare rock exoplanets would depend on the balance of the resurfacing rate (by small craters, lava flows, etc.) and the rate of weathering by micrometeorites and the solar wind (deflected by the planetary magnetic field). On rocky exoplanets, a very small residual atmosphere would be sufficient to prevent space weathering, even if that atmosphere was too thin to be detectable in transit. If rocky exoplanets have plate tectonics, then continued volcanism would reset the darkening caused by space weathering \citep{Heck2011,Foley2012}.

\subsubsection{Brightening Processes}
\label{sec:jedi}

\begin{figure}
    \centering
    \includegraphics[width=\linewidth]{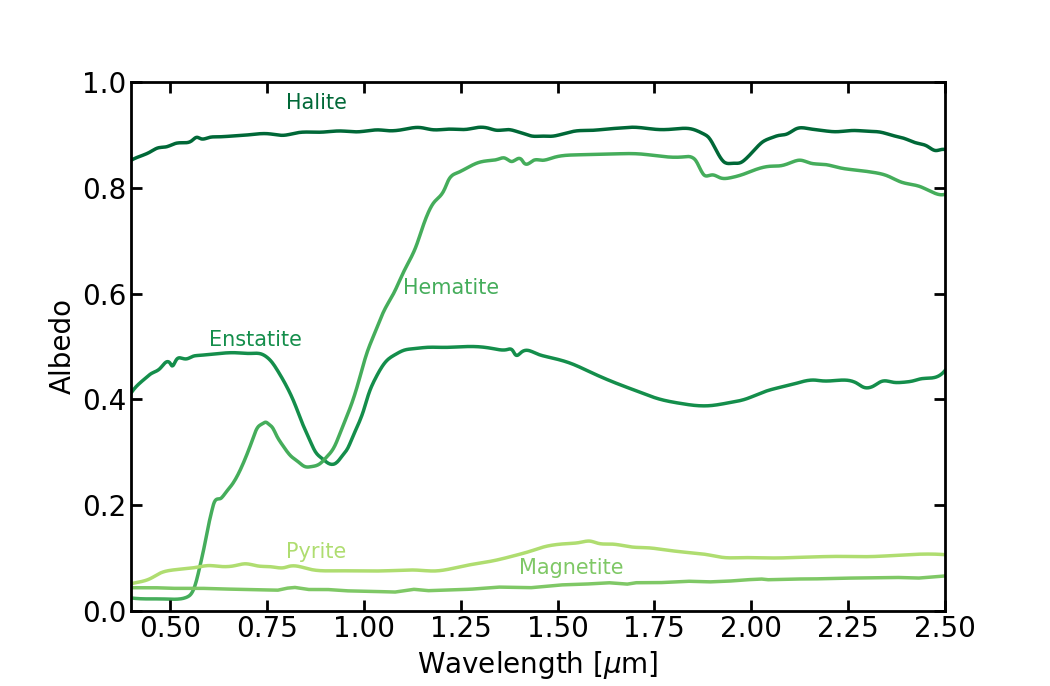}
    \caption{Near infrared reflectance spectra of minerals that might hypothetically form planet surfaces, but are not observed to be the primary material on the surface of any large solar system bodies \citep{Grove1992}. We do not consider these surfaces to be likely for worlds in $T_{sub} =$ 420-1250K orbits, for the reasons given in Section 4.2.}
    \label{fig:jedisurfaces}
\end{figure}

In spite of the reasoning above, what could nevertheless give a surface a high albedo in a $T=$~410-1250~K orbit? The below possibilities are described in order of how concerning they are for our proposed screening tool, and their reflectance spectra are shown in Figure~\ref{fig:jedisurfaces}.

(1) \emph{Shiny evaporated cores:} The silicate cores of low-density sub-Neptunes have T $\gtrsim$~3000~K \citep{HoweBurrows2015,Vazan2018,Bodenheimer2018}, and may have non-negligible partial pressures of ``rock" in the H$_2$-rich envelope, especially during the first $\sim$1\% of the planet's lifetime \citep{Fegley2016,Brouwers2018}. If an H$_2$-rich envelope of 0.1-1 wt\% of planet mass originally exists, but is subsequently stripped away, it will form a super-Earth that is an  ``evaporated core'' \citep{Owen2019}. During this evaporation process, volatile and low-molecular-weight ``rock" species will join the gas outflow \citep{Hunten1987}. The core-envelope interface will cool because the H$_2$-induced warming goes down. The equilibrium vapor pressures of ``rock'' in the envelope will also go down, and so less-volatile species will condense at the core-envelope interface and be stirred back into the liquid silicate. When the core-envelope interface cools below $\sim$1673~K the liquid-silicate stirring will stop and anything still dissolved in the envelope will form onion-shell layers at the surface as it condenses. It is conceivable that the outer shell layers would have high albedo. Na is the best candidate among the major rock-forming elements for a species that is volatile enough to have a non-negligible saturation vapor pressure at $\sim$1673~K but has a high enough molecular weight and is refractory enough that it need not escape with the hydrogen \citep{Schaefer2009}. However, creating a Na-metal surface through this scenario may require fine-tuning for an XUV flux intense and prolonged enough to shed all of the H$_2$ but weak enough that the other gases are not entrained away with the H$_2$.

(2) \emph{Reflective metals/sulfides:} Metals volatilized during an H$_2$O-rich or CO$_2$-rich atmosphere phase could add a reflective coat to the surface. This has been proposed for the highlands of Venus \citep[e.g.,][]{SchaeferFegley2004}. However, Figure~\ref{fig:jedisurfaces} shows that pyrite, the most common sulfide among Earth minerals, has a relatively low albedo, and so would be distinguishable from an atmosphere using our method.

(3) \emph{Iron oxides:} If the planet orbits an M-star, there could be significant oxidation due to the photodissociation of H$_{2}$O and the escape of hydrogen. In this case, we could observe a surface covered in iron oxides, some of which are very reflective in the near infrared. Such surfaces could have even greater Fe-oxide abundances than the case considered by \citet{Hu2012} (50\% nanophase hematite, 50\% basalt) and shown in Fig. \ref{fig:albedos} (referred to as ``Fe-oxidized''). Figure~\ref{fig:jedisurfaces} shows two examples of common iron oxides, hematite and magnetite, which span the range of reflectance spectra of iron oxides. While hematite has a high albedo at wavelengths longer than $\approx1.25$~$\mu$m, its albedo is relatively low at shorter wavelengths where the stars we consider emit the majority of their light.

(4) \emph{Pure iron:} If all of the rock has been removed we could observe a bare iron surface. A completely iron surface is unlikely according to models of collisional mantle stripping \citep{Marcus2010}, but this could be tested by using radial velocity measurements to constrain the planet's mass.

(5) \emph{Salt flats:} The early escape of a steam atmosphere might lead to a surface covered in high-albedo salt flats. Figure~\ref{fig:jedisurfaces} shows that the salt halite has a high albedo. However, volcanism on the planet would likely lead to these primordial salt flats being buried by low-albedo lava.

In addition to these possibilities, there is a wide variety of possible surface minerals \citep[e.g.,][]{Clark2007} that are not included in the surface types investigated by \citet{Hu2012}. However, we struggle to come up with plausible scenarios that would result in the bulk of a planetary surface made up of other minerals not considered here, so for this paper we focus on the well-characterized surfaces that are known to exist on planetary bodies in the solar system.

\subsection{Relationship to Other Methods of Atmospheric Detection}

\label{sec:familytree}

\begin{figure}
    \centering
    \includegraphics[width=\linewidth]{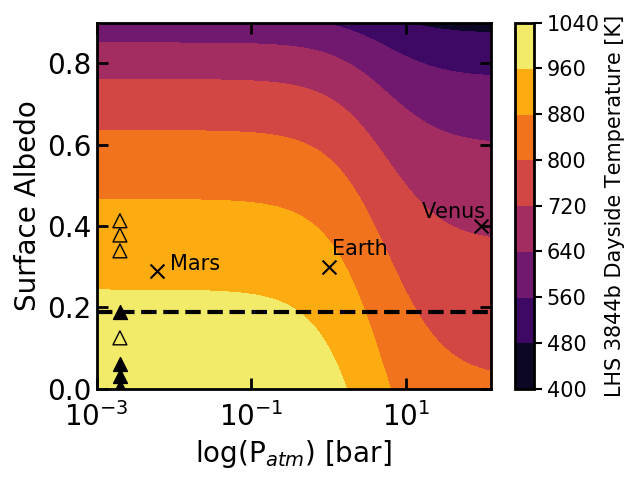}
    \caption{Relationship between the atmospheric detection method of \citet{Koll2019} and the method we present in this paper. Colored contours indicate planet dayside temperatures for LHS~3844b for a variety of atmospheric pressures and surface albedos, calculated following \citet{Koll2019}. Black $\times$ marks show the location of the rocky solar system planets on this plot. Filled triangles indicate the albedos of geologically plausible surfaces (e.g., basaltic or ultramafic) and empty triangles indicate the albedos of surfaces that are geologically implausible for $T_{sub}=410$~--~$1250$~K worlds (e.g., granitoid or ice-rich). The method of \citet{Koll2019} would have the necessary sensitivity to detect an atmosphere which changes the dayside temperature by greater than or equal to a certain amount (above/to the right of a given temperature contour). Our method allows for detection of an atmosphere at any pressure with an albedo higher than that of the most reflective plausible surface (above the dashed line).}
    \label{fig:complementary}
\end{figure}

Our method of using albedo to test for the presence of an atmosphere is complementary to that of \citet{Koll2019}, who consider the possibility of detecting an atmosphere through measurements of reduced dayside thermal emission or heat redistribution. Figure~\ref{fig:complementary} shows the relationship between these two methods of atmospheric detection. Colored contours on this plot indicate dayside effective temperatures for LHS~3844b for a variety of atmospheric pressures and surface albedos, while triangles indicate the albedos of surfaces we consider in this paper. \citet{Koll2019} present a method to detect an atmosphere that is thick enough to change the dayside temperature by greater than or equal to a certain amount (above/to the right of a given temperature contour in Figure~\ref{fig:complementary}), while our method allows detection of an atmosphere with an albedo higher than that of the most reflective plausible surface (above the dashed horizontal line in Figure~\ref{fig:complementary}).

While \citet{Koll2019} find that an atmosphere thicker than about 1 bar will transport enough heat that its secondary eclipse depth will deviate from that of a bare rock, there are several ways to create a thinner atmosphere that has a high albedo. Any planet with plentiful surface condensables can make optically thick clouds at pressures well below those needed to shift heat between hemispheres. For example, Mars has a surface pressure of 6~mbar, but has CO$_{2}$ clouds, H$_{2}$O clouds, and dust storms, all of which can be optically thick at both visible and infrared wavelengths \citep{Smith2008,Clancy2017,Haberle2017}. The top of the H$_{2}$SO$_{4}$ cloud deck on Venus is at a pressure level of $\approx1$~bar \citep{Esposito1983}. Triton's clouds have an optical depth $>$0.1, and it is plausible that under a slightly different insolation Triton could make optically thick high-albedo clouds. Finally, sulfur hazes derived from volcanic sulfur can also be very reflective \citep{Gao2017}.

\section{Conclusions}

\label{sec:conclude}

We present a method to distinguish a hot rocky exoplanet without an atmosphere from one that retains an atmosphere through measuring the planet's Bond albedo. This method is complementary to other proposed methods of atmosphere detection, including through transit or eclipse spectroscopy, reduced phase curve amplitude, or reduced secondary eclipse depth \citep{Seager2009,Morley2017,Koll2019}. Our method allows the detection of an atmosphere that is too thin to transport enough heat to impact the secondary eclipse depth but is thick enough to support high-albedo clouds.

We find that this method can be used effectively for planets with substellar temperatures of $T_{sub}=$410-1250~K. At lower temperatures, high-albedo surfaces associated with water can exist and may complicate the interpretation of a high-albedo detection. At higher temperatures, partial devolatilization of the rock may produce a high-albedo patch at the substellar point.

We investigate the properties of eight plausible surface compositions \citep{Hu2012}. We determine that an ultramafic surface is the highest-albedo ($\alpha \approx 0.19$) surface that would be likely to exist in a $T=$~410-1250~K orbit. For this surface (and the other surfaces investigated), cloud layers with optical depths of $\tau>$~0.5~--~7 will have high enough albedos to be distinguished from a bare rock surface.

\acknowledgements
We thank the anonymous reviewer, whose comments greatly improved the paper. We thank R. Milliken, P. Gao, M. Hirschmann, P. Boehnke, D.J. Stevenson, and E. Gaidos for discussions. This work was supported in part by NASA grant NNX16AB44G to E.\,S.\,K. D.\,D.\,B.\,K. was supported by a James McDonnell Foundation postdoctoral fellowship. M.\,M. acknowledges support from the Swiss National Science Foundation under the Early Postdoc Mobility grant P2BEP2\_181705.

\end{document}